\documentclass[a4paper]{cas-sc}
\usepackage{ulem}
\usepackage{array}
\usepackage[sort&compress, numbers]{natbib}
\usepackage{makecell}
\usepackage{longtable}
\usepackage{xcolor}
\usepackage{enumitem}
\usepackage{ltablex}
\usepackage{amsmath}
\usepackage{graphicx}
\usepackage{lineno}
\usepackage{longtable}
\usepackage{rotating}
\usepackage{xspace}
\usepackage{blindtext}
\usepackage{todonotes}
\usepackage{subfigure}
\usepackage{tcolorbox}
\usepackage{multirow}
\usepackage[title,toc,titletoc,header]{appendix}
\usepackage{supertabular}
\usepackage{color}
\usepackage{breakcites}
\usepackage[justification=centering]{caption}

\def\tsc#1{\csdef{#1}{\textsc{\lowercase{#1}}\xspace}}
\tsc{WGM}
\tsc{QE}
\tsc{EP}
\tsc{PMS}
\tsc{BEC}
\tsc{DE}

\begin{document}
\let\WriteBookmarks\relax
\def\floatpagepagefraction{1}
\def\textpagefraction{.001}
\shorttitle{Can we Benchmark Code Review Studies?}
\shortauthors{D.Wang et~al.}
\title [mode = title]{Can We Benchmark Code Review Studies? A Systematic Mapping Study of Methodology, Dataset, and Metric}

\author[]{Dong Wang*}[type=,
                        auid=,bioid=,
                        prefix=,
                        role=,
                        orcid=]
% \ead{}
% \ead[url]{}
\author[]{Yuki Ueda}[type=,
                        auid=,bioid=,
                        prefix=,
                        role=,
                        orcid=]
\credit{Conceptualization of this study, Methodology, Software}
\address[NAIST]{Nara Institute of Science and Technology, Japan}
\author{Raula Gaikovina Kula}
\credit{Data curation, Writing - Original draft preparation}
\author{Takashi Ishio}
\author{Kenichi Matsumoto}

\newcommand{\commentt}[1]{\indent\indent\hangindent=2.5em{\color{darkblue}#1}}

\newcommand{\RqOne}{(RQ1): \emph{What contributions and methodologies does CR research target?}\xspace}
\newcommand\respond[1]{{{#1}}}

\newcommand\nnfootnote[1]{%
  \begin{NoHyper}
  \renewcommand\thefootnote{}\footnote{#1}%
  \addtocounter{footnote}{-1}%
  \end{NoHyper}
}

\newcommand\wang[1]{\textcolor{red}{{\it [#1]}}}
\newcommand{\revise}[1]{\textcolor{purple}{\textbf{#1}}}
\newcommand\raula[1]{{\textcolor{red}{\textbf{RAULA:}#1}}}
\newcommand\ueda[1]{{\textcolor{blue}{\textbf{UEDA:}#1}}}
\newcommand\ishio[1]{{\textcolor{green}{\textbf{ISHIO:}#1}}}

\newcommand{\RqTwo}{(RQ2): \emph{How much CR research has the potential for replicability?}\xspace}
    
\newcommand{\RqThree}{(RQ3): \emph{What metrics and topics are used with CR studies?}\xspace}
	
\newcommand{\RqFour}{(RQ4): \emph{Which SE locations include papers on CR research?}\xspace}

\begin{abstract}
\noindent\respond{\textit{Context:} Code Review (CR) is the cornerstone for software quality assurance and a crucial practice for software development. As CR research matures, it can be difficult to keep track of the best practices and state-of-the-art in methodology, dataset, and metric.}

\noindent\respond{\textit{Objective:} This paper investigates the potential of benchmarking by collecting methodology, dataset, and metric of CR studies.}

\noindent\respond{\textit{Method:} A systematic mapping study was conducted. A total of 112 studies from 19,847 papers published in high-impact venues between the years 2011 and 2019 were selected and analyzed.}

\noindent\respond{\textit{Results:} First, we find that empirical evaluation is the most common methodology (65\% of papers), with solution and experience being the least common methodology. Second, we highlight 50\% of papers that use the quantitative method or mixed-method have the potential for replicability. Third, we identify 457 metrics that are grouped into sixteen core metric sets, applied to nine Software Engineering topics, showing different research topics tend to use specific metric sets.}

\noindent\respond{\textit{Conclusion:} We conclude that at this stage, we cannot benchmark CR studies. Nevertheless, a common benchmark will facilitate new researchers, including experts from other fields, to innovate new techniques and build on top of already established methodologies. A full replication is available at {\url{https://naist-se.github.io/code-review/}}.}
\end{abstract}


\maketitle
\section{Introduction}
\label{intro}
Code Review (CR) has always been the cornerstone for software quality assurance and is a crucial practice for software development teams. 
CR not only has the benefits of finding defects, but assists with other activities such as knowledge transfer and team awareness within software teams~\citep{p18}.
Microsoft reveals how \textit{``CRs at Microsoft are an integral part of the development process that thousands of engineers perceive it as a great best practice and most high-performing teams spend a lot of time doing''.}\footnote{\url{https://www.codegrip.tech/productivity/how-microsoft-does-its-code-review/}}
The rise of contemporary review tools has brought the availability of data, with the review process now being light-weight.
Contemporary tool-based reviews (such as Gerrit\footnote{\url{https://www.Gerritcodereview.com/}}, Codestriker\footnote{\url{http://codestriker.sourceforge.net/}}, and ReviewBoard\footnote{\url{https://www.reviewboard.org/}}) are widely used in both open source and proprietary software projects.
As CR research increases, so does the diversity of research methodologies, datasets, and metrics also increases, making it difficult to keep track of best practices.

This paper collects methodology, dataset, and metric for CR studies, with the end-goal to investigate the potential of benchmarking.
Other fields, such as bio-medicine\footnote{\url{https://www.biomedcentral.com/collections/benchmarkingstudies}}, 
use benchmarking studies to address the issue of:
    \textit{`as increasing numbers of methods are published in certain fields, it can be difficult to keep track of best practices for their use'.}
Large-scale studies that benchmark these methodologies on a wide range of datasets can be extremely useful to the scientific community.
In this regard, we conduct a systematic mapping study\respond{, executing the guidelines provided by \citet{Feldt_08}}.
The scope of our systematic study revolves around three research questions: \respond{to uncover (RQ1)} the state of {{contributions} and {methodologies} for research}, (RQ2) {{replicability} of existing research}, and (RQ3) {the metrics used in CR research}.

Through a collection of 19,847 papers from the high-impact SE venues, we generate visual maps for the 112 collected papers including 80 conferences and 32 journals. 
\respond{For \textit{RQ1}, we find that evaluation is the most common methodology (i.e., 73 papers), targeting particularly socio-technical and the understanding aspects of the CR process. However, there is a lack of papers that report the experience and propose solutions to deal with CR problems (i.e., four papers and thirteen papers, respectively).}
\respond{For \textit{RQ2}, our results show that CR research not only relies on the data sources from the CR process but also largely uses the data sources from the software development process (i.e., issue tracking system and GitHub).
We observe that 50\% of researches provide replicable datasets, i.e., 42 papers out of 84 papers that use quantitative or mixed-method.}
\respond{For \textit{RQ3}, we grouped 457 metrics that are used in the quantitative research into sixteen core sets (i.e., \textit{experience, code, ownership, comment, file, participant, temporal, revision, description, module, defect, queue, workload, decision, language, log, and others}) and classified nine research topics (i.e., \textit{Quality Assurance, Review Process Prediction, Acceptance Predication, Review Process, Review Participation, Review Process Prediction, Review Process Comments, CI \& Review, Test \& Review and Technical/Non-technical \& Review}). We observe that the SE topic of quality assurance is more likely to use metrics to conduct the research, with thirteen papers being studied. In addition, our mapping shows that experience and code metric sets are the two most frequent metrics used in the quantitative study.
From the mapping between metric sets and research topics, we find that different research topics tend to use particular metric sets.}

\respond{Upon further mapping between the common datasets and the metrics, we conclude that at this stage, a benchmark for CR studies is not mature but has a much-needed potential.
The map shows that papers prefer to construct their own metrics and datasets.
To promote the common dataset usage, we encourage the future researches to strive for a replicable dataset.
With the rise of machine learning and AI techniques, 
CR researchers will soon need to evaluate performance accurately against a state-of-the-art benchmark.
We envision that such a benchmark will facilitate new researchers, including experts from other fields, to \respond{propose} new techniques and build on top of already established methodologies.}

\respond{To highlight the novelty of our mapping study, we did a comparative analysis of existing systematic reviews in CR, following the protocol provided by~\citet{Feldt_08}. 
Three secondary studies~\citep{2008_mapping,short_mapping,Iwor} are identified before or in 2019. \citet{2008_mapping} conducted a mapping study on the software inspections, which is outdated, not targeting the tool-based reviews. \citet{Iwor} focused on the specific theme of tool-based code review, i.e., refactoring-awareness. \citet{short_mapping} did a preliminary study on understanding the research topic evolution of the tool-based review field. Although these two systematic studies are highly relevant to the field, our study first gives a visual summary of the potential of the benchmark with in-depth analysis. Specifically, we only study those papers that are published in the premium venues, as we believe that high-quality papers are deemed to form a best representative view for future researches. Furthermore, the outcome of this mapping study is a listing of methodologies and contributions, 42 available datasets, and 457 metrics.}

The remainder of this paper is organized as follows.
Section 2 presents the systematic mapping process, including research questions, search conduction, screening process, classification schemes, and data extraction. 
Section 3 shows the results of the systematic mapping study. 
Section 4 describes the comparative analysis of existing systematic reviews relevant to CR.
Section 5 discloses the challenges for the benchmark of dataset and metric.
Section 6 explains the threats to the validity of the research.
Finally, we summarize this paper in Section 7.

\section{The Systematic Mapping Process}
Our process is based on the work of \citet{Feldt_08} and similar to the systematic mapping study performed by \citet{mci/Abelein2016}.
Essentially, our process steps of the systematic mapping study are the definition of the research questions, search conduction of papers, screening process, keywording for the mapping, and data extraction. 

\subsection{Research Questions}
To define the scope of the mapping study, we formulate the following research questions:

\begin{enumerate}
\item \textit{\RqOne} 
The motivation for the first research question is to understand the current focus of research.
Based on the work of \citet{p18}, we would like to map out the outcomes, expectations, and contributions that the most impactful CR research tackles from the point of view of both a practitioner and researcher.

\item \textit{{\RqTwo}}
The motivation for the second research question is to understand how the data source impacts CR research.
Understanding the sources can provide insight into the current state and gaps in terms of the data collection and availability.
Furthermore, there has been growing initiatives to make data open and replicability which is encouraged in the community~\footnote{A recent initiative by the Springer EMSE Journal shows how research is working towards open science and replicable studies at \url{https://github.com/emsejournal/openscience}}.
\item \textit{{\RqThree}} The motivation for the third research question is to uncover what kinds of metrics are used in CR empirical studies. Understanding how the metrics are used and the associated topics can motivate the potential to benchmark CR studies.
\end{enumerate}

\begin{figure}[pos = t]
\centering
\includegraphics[width=0.48\textwidth]{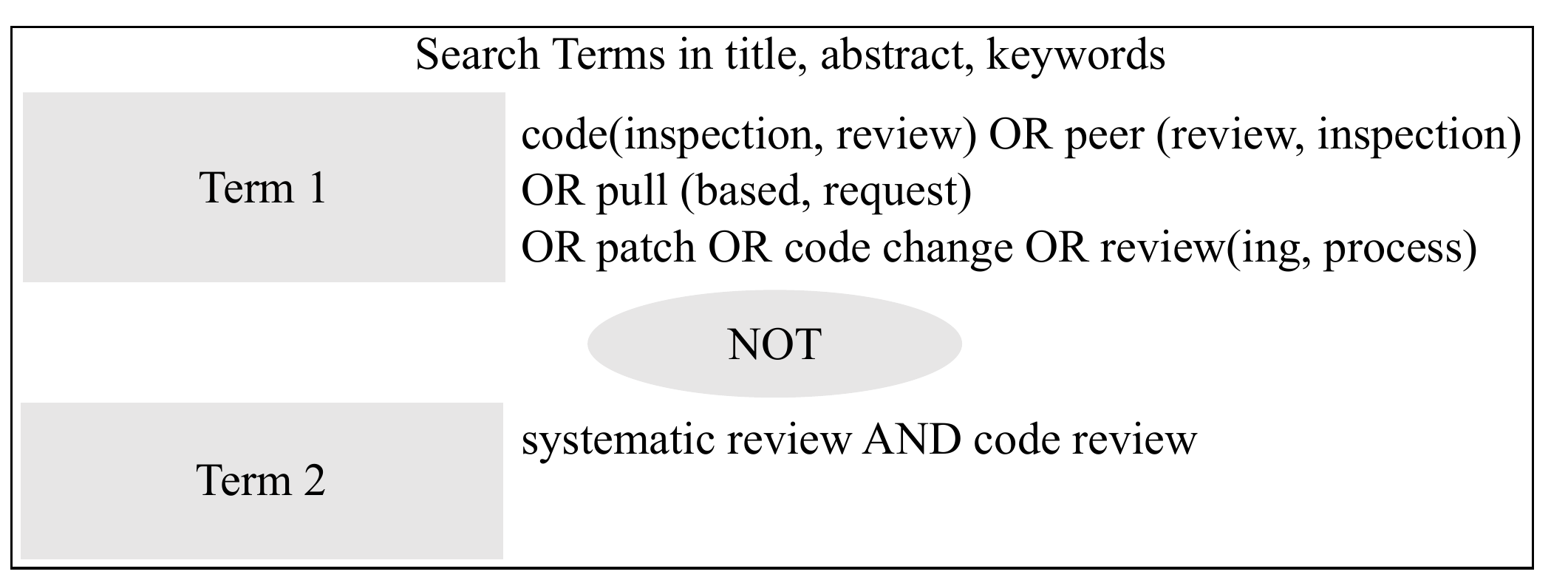}
\caption{Defined terms used in the search strings}
\label{fig:search}
\end{figure}
%%%%%%%%%%%%%%%%%%%%%%% 
\begin{table}[pos = b]
\caption[Caption for LOF]{Corpus of venues (conferences and journals) studied in this paper. 
Note that ICSM now is called ICSME; and WCRE and CSMR are fused into SANER. In addition, the GPCE h5-index is not retrieved but is ranked as B.}
\resizebox{\textwidth}{!}{
\begin{tabular}{llll}
Journal  & Name                                                                             & Impact factor & Established \\ \hline\hline
            TSE    & IEEE Transactions on Software Engineering                                        & 6.112 & 1991  \\
                        ESE    & Empirical Software Engineering                                                   & 3.156 & 1996  \\
            IST    & Information and Software Technology                                              & 2.726 & 1992  \\
            S/W    & IEEE Software                                                                    & 2.589 & 1991  \\
            TOSEM  & Transactions on Software Engineering and Methodology                             & 2.516 & 1992  \\
             JSS    & The Journal of Systems and Software                                              & 2.450 & 1991  \\
            REJ    & Requirements Engineering Journal                                                 &  1.933  & 1996  \\
 SOSYM  & Software and System Modeling                                                     & 1.876 & 2002  \\
            ASEJ   & Automated Software Engineering Journal                                           & 1.857   & 1994  \\
            SPE    & Software: Practice and Experience                                                &  1.786  & 1991  \\
            SQJ    & Software Quality Journal                                                         & 1.460   & 1995  \\
            STVR   & Software Testing, Verification and Reliability                                   &  1.226  & 1992  \\
            SMR    & Journal of Software: Evolution and Process                                       & 1.178   & 1991  \\
            ISSE   & Innovations in Systems and Software Engineering                                  & 0.950   & 2005  \\
            IJSEKE & International Journal of Software Engineering and Knowledge Engineering          & 0.886   & 1991  \\
            NOTES  & ACM SIGSOFT Software Engineering Notes                                           &  0.490   & 1999  \\\hline
            Conference  & Name                                                                             & h5-index & Established \\ \hline\hline
            ICSE   & International Conference on Software Engineering                                 & 75 & 1994  \\
             FSE    & ACM SIGSOFT Symposium on the Foundations of Software Engineering                 & 51 & 1993  \\
            ASE    & IEEE/ACM International Conference on Automated Software Engineering              & 40 & 1994  \\
            MSR    & Working Conference on Mining Software Repositories                               & 38 & 2004  \\
            ISSTA  & International Symposium on Software Testing and Analysis                         & 35 & 1989  \\
            ICSM   & IEEE International Conference on Software Maintenance                            & 33 & 1994  \\
            ICPC   & IEEE International Conference on Program Comprehension                           & 33 & 1997  \\
            SANER  & IEEE International Conference on Software Analysis, Evolution and Re-engineering & 30 & 2014  \\
             ICST   & IEEE International Conference on Software Testing, Verification and Validation   & 27 & 2008  \\
            RE     & IEEE International Requirements Engineering Conference                           & 25 & 1993  \\
            CSMR   & European Conference on Software Maintenance and Re-engineering                   & 25 & 1997  \\
             WCRE   & Working Conference on Reverse Engineering                                        & 22 & 1995  \\
  MDLS   & International Conference On Model Driven Engineering Languages And Systems       & 21 & 2005  \\
            ESEM   & International Symposium on Empirical Software Engineering and Measurement        & 20 & 2007  \\
            FASE   & International Conference on Fundamental Approaches to Software Engineering       & 18 & 1998  \\
            SSBSE  & International Symposium on Search Based Software Engineering                     & 15 & 2011  \\
            SCAM   & International Working Conference on Source Code Analysis \& Manipulation         & 15 & 2001  \\
            GPCE   & Generative Programming and Component Engineering                                 &    & 2000  \\ \hline
\end{tabular}}
\label{paper_source}
\end{table}

\subsection{Conduct Search}
We use the following strict characteristics as recommended by \citet{Kitchenham_07} to formulate our search string: (C1) a defined search strategy, (C2) a defined search string, based on a list of synonyms combined by ANDs and ORs, (C3) a broad collection of search sources, (C4) strict documentation of the search, (C5) paper selection should be checked by at least two researchers.
Figure \ref{fig:search} shows the defined search string.
\respond{For Term 1 in the search string, as shown in Figure \ref{fig:search}, We apply commonly used terminologies in CR to search, such as code inspection, code review, peer review, peer inspection, or tools such as pull requests or pull-based. 
To enlarge the paper dataset, other terminologies such as patch, code changes, and reviewing are also considered as appropriate for CR research.}
\respond{For Term 2 in the search string, we do not include the CR related papers that are in the form of the systematic review.
Nevertheless, we separately conduct a comparative analysis of existing systematic reviews in Section 4.}

Based on RQ1, 
to ensure papers of high quality and understand the state-of-the-art in the field, we specifically searched for papers published in the high-impact journals and conferences from the software engineering domain \respond{between 2011 and 2019}.
\respond{Inspired by the work of \citet{short_mapping}, we select the 2011 time-frame as our starting point, since Badampudi et al. identified a steady upward trend regarding publications related to the contemporary tool-based review starting in 2011. In addition, several well-known open-source projects (e.g., OpenStack and Qt projects) use the Gerrit platform since 2011.}
\respond{Table \ref{paper_source} shows the summary of paper collection source. Regarding the publication venue selection, similar to the mapping study conducted by~\citet{topic_tse},} papers were extracted from 18 conferences (i.e., International Conference on Software Engineering) with relatively high h5-index and
16 journals with high impact factors.
\respond{h5 is the
h-index for articles published in a period of 5 complete years obtained from Google Scholar. We rely on Guide2Research\footnote{\url{https://www.guide2research.com/}} to retrieve the h5-index. Although Generative Programming and Component Engineering (GPCE) is not recorded with the h5-index, it is ranked as B according to core ranking.\footnote{\url{http://portal.core.edu.au/conf-ranks/}} The Impact Factor (IF) is is an index (numerical value) to evaluate how much impact a journal (scientific journals) has, based on Clarivate Analytics.\footnote{\url{https://clarivate.com/webofsciencegroup/essays/impact-factor/}} The conferences with higher h5-index or the journals with higher impact factors are deemed to carry more intrinsic prestige in their respective fields.}
To reduce its selection bias, we selected from a wide range of digital resources to follow (C3) a broad collection of search sources: ACM Digital Library, IEEE Xplore, Science Direct, and SpringerLink databases.
For example, the data from 2012 to 2019 for Mining Software Repositories Conference is collected through IEEE Xplore, but the data from 2011 is available in ACM Digital Library.
We extracted 19,847 papers from the above four search sources that were published in the last nine years (i.e., 2011$\sim$2019), as shown in Table~\ref{tab:collected}.

\respond{Assessing the quality of primary papers can be used as an additional criterion for the exclusion~\citep{procedure_2004}. As part of our quality assessment, we exclusively consider papers from these premium venues as we assume they are of high quality and widely get recognized within the SE domain. Additionally, to ensure only technical contributions, in the further data processing, we filter out short papers, editorials, tutorials, panels, poster sessions and prefaces, and opinions (8 pages or less). Nonetheless, internal and conclusion threats may exist,  and we further discuss the threats with regard to this assessment in Section 6. After the conduct search, we were able to get 437 initial papers, as shown in Table \ref{tab:collected}.}

\subsection{Screening Process}
Our screening process is comprised of inclusion and exclusion criteria.
For this manual exclusion, the following inclusion and exclusion criteria were applied to the abstract of each paper.
\textit{Inclusion criteria:} \respond{Three inclusion criteria are defined}, namely,
\textit{(IC1)}: paper should focus on topics on code inspections, code review, code review tools, pull request,
\respond{\textit{(IC2)}: the paper is peer reviewed,}
\respond{\textit{(IC3)}: the paper is written in English and the paper has full text available.}
\textit{Exclusion criteria:} Four exclusion criteria were defined that cover the datasets, purposes and the evaluation of the studies. 
The following papers were excluded that met these criteria: \textit{(EC1)}: the paper does not mention any CR activities, \textit{(EC2)}: the paper focuses on other software development process, e.g., issue tracking process, continuous integration, testing, \textit{(EC3)}: the paper is out of scope with focusing on other sub-fields such as program analysis, code clone, defect prediction, refactoring, social technique,
\respond{\textit{(EC4)}: the paper is outsides our studied time-frame.}

%%%%%%%%%%%%%%%%%%%%%%%%%%%%%%%%%%%%%%%
\begin{table}[pos = hpt]
\caption{Statistics of the filtering of the papers during the conduct search and screening process}
\centering
\begin{tabular}{llrr}
&&\# of Papers\\
\hline
Conduct Search&&\\
&Search String Result& 19,847\\
\textit{All Papers}&&& 437\\
\hline
Screening of Papers&\\
%&IC and EC (Premium)& 62\\
&Conference paper & 80\\
&Journal paper & 32\\
\textit{Total Papers}&&& 112\\\hline
\label{tab:collected}
\end{tabular}
\end{table}
%%%%%%%%%%%%%%%%%%%%%%%%%%%%%%%%%%%%%%%%%

\begin{figure}[pos = t]
\centering   
\subfigure[Distribution of paper publication\label{preliminary_a}]{\includegraphics[width=70mm,height=40mm]{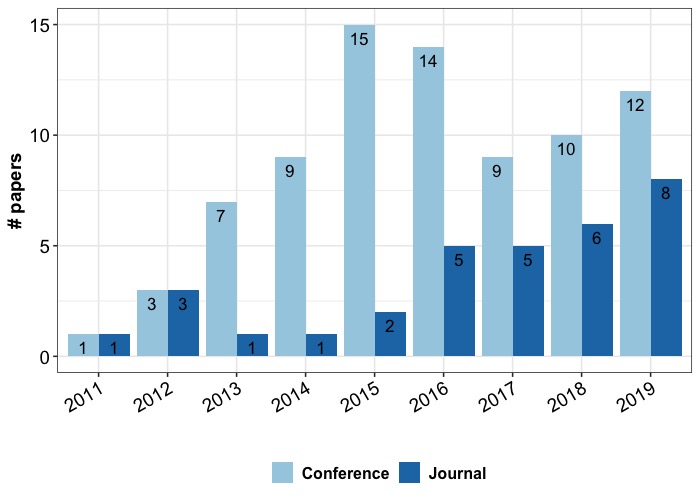}}
\subfigure[Distribution of research types \label{preliminary_b}]{\includegraphics[width=70mm,height=40mm]{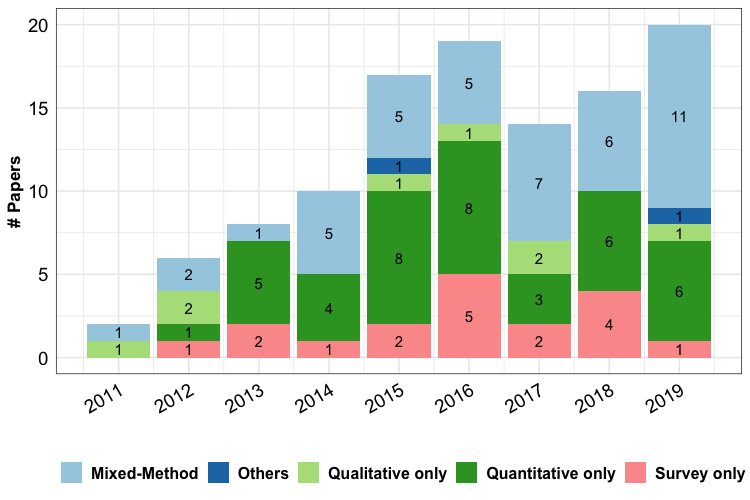}}
\caption{The distribution of paper publication and their research types yearly from 2011 to 2019. CR papers keep an upward trend and the journal becomes a popular choice for publication. Mixed-Method papers becomes popular in the recent time.}
\label{preliminary}
\end{figure}

To reduce bias and follow (C5), this manual paper selection was conducted by the first and the second authors.
As a result of the screening process, we were able to collect 112 papers out of 437 initial papers, which include 80 premium conference papers and 32 high-impact journal papers, as shown in Table~\ref{tab:collected}.
Figure~\ref{preliminary_a} depicts the distribution of these 112 papers based on conferences and journals during our studied time frame.
\respond{In detail, the figure shows that the CR research publications keep an upward trend in the recent three years, i.e., fourteen, sixteen, and twenty papers are published in 2017, 2018, and 2019.}
We also observed that papers submitted to journals are on the upward trend from 2015, i.e., eight papers were submitted to journals in 2019.

\respond{To further explore the trend of the research type, we manually classified the types of research papers (\respond{e.g.,} quantitative and qualitative) according to the work of \citet{approach}.
We classify the research types into four categories: i) Quantitative only, ii) Quantitative only, iii) Mixed-Method, and iv) Survey only. 
The Mixed-Method refers to those papers using the combination of quantitative method and qualitative/survey method.
For the Survey only type, it not only includes survey but also includes interview and user/control studies.
We classify papers which do not fit the above types into others.
We classify research paper types with two rounds.
First, two authors classified them in the first round.
In the second round, the third author full with research experience joined to validate each collected paper. Figure~\ref{preliminary_b} shows the research type distribution of 112 papers during our studied time frame. We observe that the Mixed-Method papers become popular in the recent time, i.e., eleven Mixed-method papers are published in 2019.}

\begin{table}[pos =hpt]
\caption{Summary of the classification scheme used to identify contribution, methodology, replication, and metric.}
\label{tab:datasummary}
\centering
\resizebox{\textwidth}{!}{
\begin{tabular}{m{2em}|c|l|p{0.55\linewidth}}
\hline
Class & Sub-class & Category & Description \\\hline\hline
\multirow{10}{*}{\rotatebox{90}{Contributions}} & Practitioner & Communication \cite{p18} & The developers are provided with the need of richer communication than comments annotating the changed code when reviewing. 
Teams should provide mechanisms for in-person or, at least, synchronous communication. \\\cline{3-4}
& & Potential Benefit \cite{p18} & Modern CR provides benefits beyond finding defects. 
CR can be used to improve code style, find alternative solutions, increase learning, share code ownership, etc. 
This should guide CR policies. \\\cline{3-4}
& & Quality Assurance \cite{p18} & CR does not result in identifying defects as often as project members would like and even more rarely detects deep, subtle, or “macro” level issues.\\\cline{3-4}
& & Understanding \cite{p18} & When reviewers have prior knowledge of the context and the code, they complete reviews more quickly and provide more valuable feedback to author.\\\cline{2-4}
& Researcher &Automation \cite{p18}& Tools for enforcing team code conventions, checking for typos, and identifying dead code already exist.
Even more advanced tasks such as checking boundary conditions or catching common mistakes have been shown to work in practice on real code.
Automating these tasks frees reviewers to look for deeper, more subtle defects.\\\cline{3-4}
& & Program comprehension \cite{p18} & Context and change understanding are challenges that developers face when reviewing, with a direct relationship to the quality of review comments. \\\cline{3-4}
& & Socio-technical effect \cite{p18} & These are studies that involves the consideration of both human and technical aspects. In terms of CR, Studies can be designed and carried out to determine if and how team collaboration, coordination, awareness and learning occurs.\\\hline
\multirow{10}{*}{\rotatebox{90}{Methodologies}} & -- & 
Validation Research \cite{Wieringa:2005} & Techniques investigated are novel and have not yet been implemented in practice. Techniques used are for example experiments, i.e, work done in lab \\\cline{3-4}
& &Evaluation Research \cite{Wieringa:2005}& Techniques are implemented in practice and an evolution of the technique is conducted.That means, it is shown how the technique is implemented in practice (solution implementation) and what are the consequences of the implementation in terms of benefits and drawbacks (implementation evaluation). This also includes to identify problems in industry. \\\cline{3-4}
& & Solution Proposal \cite{Wieringa:2005}& A solution for a problem is proposed, the solution can be either novel or a significant extension of an existing technique. The potential benefits and the applicability of the solution is shown by small example or a good line of argumentation. \\\cline{3-4}
& & Experience Paper \cite{Wieringa:2005}& Experience papers explain on what and how something has been done in practice. It has to be the personal experience of the author\\ \cline{3-4}
& & Survey Paper & These papers are qualitative studies that use a questionnaire or interviews to evaluate some phenomena  \\ \hline
\multirow{3}{*}{\rotatebox{90}{Replication}} & - & \respond{Private Datasets} & Neither dataset nor the source code is available. The study may not be replicated. \\\cline{3-4}
& & \respond{Partial Datasets} & Part of the dataset is available. The study could not be replicated fully with partial datasets.
\\\cline{3-4}
& &\respond{Public Datasets} & Replication including either full dataset or the source code is provided via hyperlinks or paper references. The study is deemed to be replicable using provided datasets.
\\
& &  &  \\\hline
\rotatebox{90}{Metric} &-& Metric sets used in empirical studies & \respond{Metrics that are used in CR research can be classified according to the level of three aspects: product, process, and people.} \\\hline
\label{tab:class}
\end{tabular}}
\end{table}
%%%%%%%%%%%%%%%%%%%%%%%%%%%%%%%%%%%%%%%%%

\subsection{Keywording of Relevant Papers}
Inspired by \citet{Feldt_08}, we classified each paper based on the scope outlined in each research question with results shown in Table \ref{tab:datasummary}.
During the  classification, it not only includes the detailed reading of the abstract, but sometimes requires a careful reading of the whole paper itself.

\textbf{Contributions and Methodologies (RQ1).}
To classify research contributions of the papers, we base our work on the work of \citet{p18}. They classify contributions for two objectives (i.e., contributions to benefit practitioner and researcher).
For the classification process, three co-authors sat in a round-table and labeled each contribution based on seven category features shown in Table \ref{tab:datasummary}.
The process was to first read the abstract and decide the classification.
If there was a dispute, then the paper was quickly analyzed and a discussion of the paper started between the co-authors before the consensus reached.

To classify methodologies that were applied to the studies, we used existing definitions of research facets \cite{Feldt_08}.
For the classification, three co-authors sat in a round-table and labeled each methodology based on the category features.
The first keywords relating to the methodology were searched and discussed. 
Similar to the keywording of contributions, the full contents of the paper were consulted if a dispute arose among the co-authors.

\textbf{Replication (RQ2).}
To classify the replicability of papers, we identified the source of the data, whether the dataset is either available via the link or is referred to a prior dataset.
Our scope is limited to the quantitative and mixed-method papers.
Since detailed information of the dataset is not likely to be in the abstracts, co-authors were required to scan the papers to extract any online links of a dataset or a reference to an existing dataset.
Furthermore, as shown in Table \ref{tab:class}, authors classified the papers according to the nature of the studied systems (i.e., open source projects or industry).
Note that the classification is non-exclusive as some studies involved projects that were both open and closed data.

\textbf{Metrics (RQ3).}
To group the metrics used in collected papers, we only scan the papers conducted in quantitative method and mixed-method. 
For the classification, we do the following two: (1) metrics mapping research aspects and (2) metrics mapping research topics.
For the first classification, we pick up all metric description tables from papers.
We then apply open card sorting to construct a taxonomy of codes of the metrics.
In detail, following the metric descriptions, the coded metrics are merged into cohesive groups that can be represented by a similar high-level code, i.e., metric sets.
In the card sorting process, three authors sit together and sort metrics until all achieve the consensus. 
Based on prior work~\cite{xinyang_2016}, the aspects of CR research can be divided into three targets: product, process, and \respond{people}.
Following these aspects, we then classify the constructed high-level metric sets using ticks.
For the second classification, the first author classified initial research topic groups by reading the abstracts and introductions.
After that, another experienced author did the validation to assure the constructed topic groups that were distinguished.

\subsection{Data Extraction and Mapping of Studies}
Using the classification scheme, we then utilize visual mappings of the results to highlight states in the collected papers.
To identify which categories have been emphasized in past research and show possible opportunities for future work, we use three plot types to show maps (i) tables, (ii) bar plots, and (iii) bubble maps.
Once the scheme is in place, we used excel spreadsheets to store the data and applied R scripts to extract and categorize the papers.
Furthermore, we put rationales to decide why we believe each paper is categorized.
Below are the visual techniques and rationale for answering each RQ:

\textbf{Visual Map of RQ1.} To answer RQ1, we show a visual mapping of the contributions (with the researchers and practitioners separately) against the methodologies. 
We intend to find out how the methodologies influence the contributions and what is the popular combination of contributions and methodologies.
A bubble map will be used to show results.
The map should show what contributions are saturated and which perceived contributions have the potential for future work.
We will also pick up examples of each classified paper for an in-depth discussion of the maps.

\textbf{Visual Map of RQ2.} To answer RQ2, we show a visual mapping of the replicability of the collected papers.  
We intend to determine how much CR research has the potential to be replicated. 
A bar chart will be used to visualize the main results.
The map should show the proportion of how many papers can be replicated and show what forms are used to provide replication (i.e., via links or reference to the dataset).
For a deeper understanding of the data sources, we perform additional sub-classification of the source: (i) research that extracts data from pure code review tools (e.g., Gerrit tools in OSS and special review systems or tools in industry such as CodeFlow tool in Microsoft), (ii) research that extracts data that not only contains CR, but expands on other software development tools such as mailing lists, version control system, GitHub, and issue tracking system, (iii) research that extracts data from observational experiments in the form of interviews, survey, and control study.
\respond{Additionally, we classify the platforms where the available datasets are provided into four types: (i) online storage, i.e., dropbox, (ii) permanent storage, i.e., zenodo, (iii) GitHub /BitBucket, and (iv) personal or university.}

\textbf{Visual Map of RQ3.} To answer RQ3, we show a visual mapping of the metric benchmark in terms of research aspects (i.e., product, process, and human) and research topics.
We intend to formulate a systematic metric group for a future research guide and understand in which topic what kinds of metrics should be included.
Two tables will be drawn to present our benchmark details. 
The first table map should show 1) how many different metrics are used with their frequency in papers and (2) what research aspects do these metrics target.
The second table map should show different combinations of metrics are used in different research topics.

\section{Results: Maps of CR Research}
The results will answer the research questions, with the visual maps of the categories of the papers.

%%%%%%%%%%%%%%%%%%%%%%%%%%%%%%%%%%%%%
\begin{figure}[pos = t]
\centering
\includegraphics[width=.8\textwidth]{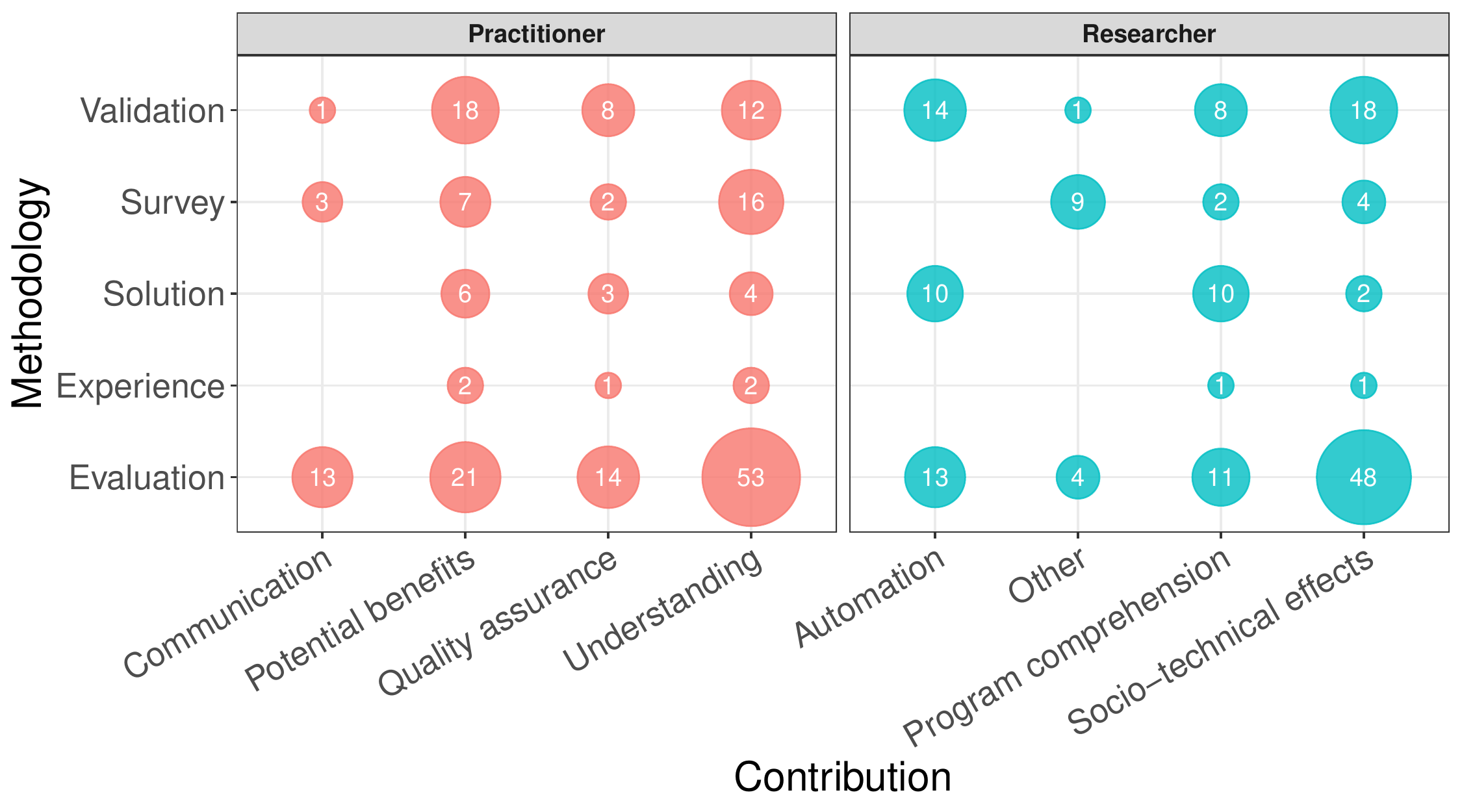}
\caption{Visual Map for RQ$_1$, showing the contribution and methodology of CR research. The figure shows that evaluation is the most popular methodology, particularly targeting the contributions of understanding and socio-technical effects.}
\label{fig:RQ1}
\end{figure}
%%%%%%%%%%%%%%%%%%%%%%%%%%%%%%%%%%

\subsection{\RqOne}

Figure \ref{fig:RQ1} shows both the saturation of papers as well as the potential research opportunities for the field. 
\respond{Note that a paper can target multiple contributions, using more than one methodology.}
The figure clearly shows that evaluation is the most popular methodology, benefiting both the practitioners and researchers.
For practitioners, most of the papers have contributions to potential benefits and understanding aspects.
Potential benefits mean that modern CR provides benefits beyond the fundamental need to find defects. 
CR is demonstrated to be useful for other tasks.
We introduce three examples in detail below.
For the task of improving code style,
\citet{p12} presented an interactive
approach named \textit{CRITICS} for inspecting systematic changes and the results show that it should improve developer productivity during this process.
For the task of increasing the learning, \citet{p07} conducted a large-scale survey to investigate work practices and challenges in pull-based development model and results show that integrator should consider several factors in their decision making.
For the task of review comments usefulness, \citet{p21} found that useful comments share more vocabulary
with the changed code, contain salient items like relevant code
elements, and their reviewers are generally more experienced.
For instance,
exploring how CR is conducted can be used for practitioners to better implement review activity and improve the review quality.
Understanding is when reviewers have prior knowledge of the context and the code, they complete reviews more quickly and provide more valuable feedback to the author.
Key examples are researches that look into the quality of review and types of defects. 
\citet{p09} provided a deep insight into how developers define review quality, what factors contribute to how they evaluate submitted code, and what challenges they face when performing review tasks.
\citet{p29} conducted a manual research to increase the understanding of practical benefits that the MCR
process produces on reviewed source code.
Their results show that types of changes due to the MCR
process in OSS are strikingly similar to those in the industry and
academic systems.

On the other hand, researcher-oriented CR studies mostly focus on socio-technical contributions.
Socio-technical related papers are studies that involve the consideration of both human and technical aspects. 
One popular topic is reviewers recommendation and many studies have been done on this topic.
\citet{p34} put textual information and file location analyses together to recommend reviewers more accurately.
\citet{p48} recommended code reviewers based on their expertise.
Apart from reviewer recommendation topic, many other human related researches have been conducted such as review participation~\cite{pj3}, evaluation of contributions~\cite{p41} and broadcast during CR process~\cite{p20}.
In terms of CR, studies can be designed and carried out to determine if and how team collaboration, coordination, awareness and learning occurs. 
In terms of opportunities, Figure \ref{fig:RQ1} highlights the lack of experience papers. 
This is crucial and shows a lack of reporting and feedback from developers.
Instead, we see that there are a stable number of survey papers.
Other notable potential methodologies are the experience and solution, which indicates that more practical tools need to be developed to help practitioners in reality.

Table~\ref{tab:combination} shows a listing of top five paper contributions with the methodologies used, illustrating how evaluation studies are dominant.
According to our results, two most popular combinations are the evaluation study that has a understanding contribution (e.g., 53 papers), then followed by the study with socio-technical effect target following the evaluation methodology (e.g., 48 papers).
For the understanding target with evaluation methodology, we introduce two representative studies. 
In the work of~\citet{p36}, they investigate the patch rejected reasons with 300 samples and formulate the practical suggestions for patch author submission.
This work helps patch authors to better understand what kinds of patch should be submitted so as to reduce the rejection chance.
Another work conducted by~\citet{p61} empirically analyze how developers document pull requests with external references using mixed-method.
Their results indicate that for developers, external resources are useful to learn something new or to solve specific problems.
One example of the evaluation study with a social-technical effect is \citet{p23}, which investigates CR practices in defective files combined with human factors (e.g., the participation of the reviewer in the process).
In detail, authors evaluate the results using a detailed empirical study of the Gerrit review system within the Qt project.
Similarly, \citet{p33} reported on a case study investigating CR quality for Mozilla and explore the relationships between the reviewers’ code inspections
and a set of factors, both personal and social.
It is interesting to note that 36 out of 53 papers, i.e., around 68\%, that contribute to better understanding also have socio-technical contributions.
For example, \citet{pj3} studied what factors influence review participation in the CR process which in turn helps practitioners understand the situation when they tend to join.
While within the validation methodology, we find the most popular combination is papers that target contributions of potential benefits with such methodology.
The majority of the validation papers are based on recommendation or prediction models.
An example of this type of paper is \citet{p11}, where authors suggest an approach of reviewer recommendation based on cross-project and technology experience.

\begin{table*}[pos= b]
    \centering
    \caption{Top 5 combination of contribution and methodology}
\resizebox{\textwidth}{!}{\begin{tabular}{l|lp{3cm}p{4cm}p{4cm}|}
\cline{2-5}
                                  &            & \multicolumn{3}{c|}{Contribution}                                      \\ \hline
\multicolumn{1}{|l|}{}            &            & Socio-technical effects & Understanding              & Potential Benefits \\
\multicolumn{1}{|l|}{Methodology} & Evaluation &   \cite{p37, pj1, p20, p34, p22, pj11, pj6, pj9, p24, pj5, p04, p29, p32, p52, p06, p23, pj13, p21, p16, pj4, pj12, p33, pj16, p47, pj2, p41, p42, p15, pj3, pj14, p03, p27, pj15, p28, p55, p56, pj22, p60, p64, p65, p68, p69, pj27, pj28, pj32, p79} & \cite{p37, p36, pj1, p20, p22, pj11, pj6, p24, pj5, p04, p29, p32, p17, p52, p06, p23, pj13, p16, pj4, p46, p44, p33, p35, p47, p49, p42, p15, pj3, pj14, p03, p27, pj15, p53, p28, p55, p58, pj20, pj22, pj24, p60, p61, p62, p64, p66, p67, p68, p69, pj26, pj28, pj30, pj31, pj32, p79} & \cite{pj12, pj1, p42, p34, p26, p17, p53, pj2, pj13, pj9, p21, pj4, pj5, p40, p56, pj25, p70, p71, p72, pj26, p80}\\
\multicolumn{1}{|l|}{} & \multicolumn{4}{l|}{}\\
\multicolumn{1}{|l|}{}           & Validation & \cite{p11, pj12, p34, pj6, p23, p06, p31, p47, pj13, pj2, p21, pj9, pj5, pj3, p28, p59, p74} &                         &  
\cite{p11, pj12, p34, p19, p48, p45, p31, p12, p39, pj13, pj2, p14, p21, pj8, pj9, pj5, pj19, p59}\\ \hline
\end{tabular}}
\label{tab:combination}
\end{table*}

Apart from the referred papers using two popular methodologies that are listed in the Table~\ref{tab:combination}, we also provide the complete paper list for those papers using survey, solution, and experience methodology.
Among these three methodologies, survey is relatively frequent with eighteen papers being retrieved~\cite{p01, p02, p07, p08, p09, p13, p18, p25, p30, p38, p43, p50, p62, p67, p68, pj7, pj18, p54}.
Thirteen papers are classified as solution~\cite{p10, p24, p11, p12, p14, p39, p45, p51, p63, p75, p76, p77, p78}.
The last experience methodology is the most rare case, only four papers being found~\cite{p10, p24, pj17, pj21}.

\begin{tcolorbox}
\textbf{Answering RQ1:}
    \respond{Our results show that 65\% of CR researches published in premium SE venues use sound evaluation methodology (i.e., 73 papers), targeting particularly socio-technical and understanding of CR processes. However, there is a lack of papers that report the experience and propose solutions to deal with CR problems (i.e., four papers and thirteen papers, separately).}
    \respond{The implication of RQ1 is that, we encourage the practitioners to more emphasize and share the experience with CR. At the same time, future research could propose more solution tool support to facilitate the developers to make the CR process more efficient.}
\end{tcolorbox}

\subsection{\RqTwo}

\begin{table}[pos = b]
\caption{Data source classification, Top-3 studied projects applying Gerrit review tools, and platform distribution. Mixed-method papers can be classified into Inter./Sur./Control Study as well.}
\begin{tabular}{llp{5cm}}
                                                            &                           & Relevant Papers                                                                                                                                                                                                                                                             \\ \hline
\multicolumn{1}{c|}{\multirow{3}{*}{}}           & CR Process                & \cite{p01, p04, p06, p10, p14, p19, p21, p23, p24, p28, p29, p31, p32, p34, p35, p39, p40, p42, p46, p48, p51, pj2, pj3, pj5, pj6, pj8, p55, p57, p58, p64, p65, p66, p69, p70, pj26, pj27, pj28, p79}                                                     \\ \cline{2-3} 
\multicolumn{1}{c|}{Data source}                                       & Software Dev. Process     & \cite{p03, p11, p15, p16, p17, p20, p22, p26, p32, p33, p35, p36, p37, p39, p42, p44, p45, p47, p48, p49, p52, p53, pj1, pj4, pj6, pj9, pj11, pj12, pj13, pj14, pj15, pj23, pj24, p59, p60, p61, p62, p67, p72, pj30, pj31, pj32, p75, p76, p77, p79, p80} \\ \cline{2-3} 
\multicolumn{1}{c|}{}                                       & Interview/Survey/Control Study & \cite{p10, p14, p05, p12, pj10, p36, pj16, p67, pj28, pj30, p73, p75, p76, p77, p80}                                                                                                                                                                       \\ \hline
\multicolumn{1}{l|}{\multirow{3}{*}{}} & Qt                        & \cite{p04, p06, p23, p31, p34, pj3, pj5, p28, pj6, p48, p64, p65, p69, pj26, pj28, p79}                                                                                                                                                                    \\ \cline{2-3} 
\multicolumn{1}{l|}{Top-3 Gerrit projects}                                       & OpenStack                 & \cite{p04, p06, p31, p34, pj2, pj3, pj6, p48, p58, p65, p70, pj26, pj27, pj28, p79}                                                                                                                                                                        \\ \cline{2-3} 
\multicolumn{1}{l|}{}                                       & Android                   & \cite{p31, p34, p40, pj3, pj8, p35, p32, p65, p69, pj27, pj28, p79}                                                                                                                                                                                        \\ \hline
\multicolumn{1}{l|}{\multirow{3}{*}{}} & GitHub/Bit Bucket                        & \cite{p52, p12, p17, p23, p28, p32, p34, p58, p62, p65, p68, pj3, pj4, pj5, pj6, pj13, pj28}  
   \\ \cline{2-3} \multicolumn{1}{l|}{\multirow{3}{*}{}}
& Online Storage & \cite{p67}
 \\ \cline{2-3} 
\multicolumn{1}{l|}{Dataset platform} & Personal/University                       & \cite{pj27, p03, p05, p21, p31, p33, p35, p40, p46, p48, p49, p60, p66, p70, p73, p79, pj9}  \\\cline{2-3} 
\multicolumn{1}{l|}{\multirow{3}{*}{}} & Permanent Storage                       & \cite{p04, p72, p76}  \\\cline{2-3} 
\multicolumn{1}{l|}{}                                       & Reference to Existing Data set & \cite{p06, p23, p42}     \\\hline
\end{tabular}
\label{tab:RQ2result}
\end{table}

In Table \ref{tab:RQ2result}, we divide all premium papers into different classification according to the definition of data sources shown in Section 2.5. 
We describe each data source in detail below.
In \textit{CR Process}, for example, \citet{pj5} conducted the research to see the impact of code reviews on software quality. 
They only focus on review process and extract the data from QT, VTK, ITK projects using review tools (e.g., Gerrit).
For \textit{Software Development Process}, for example, \citet{p33} investigated whether people and participation matter the quality of review. 
In their research, they collected data from issue tracking system (e.g., Bugzilla) which belongs to the development process.
In \textit{Interview/Survey/Control Study}, for instance, \citet{pj7} analyzed the process aspects and social dynamics of CR 
from the diverse surveys of Mircosoft and other open source projects.
In another example, \citet{p05} researched the representation of code in the brain with fMRI study.
They involved 29 participants to carry out the controlled experiment and got result feedback.
We find that \textit{code review process} related dataset is the most extracted from the well-studied Gerrit tool.
One advancement has been the release of the rest API, in which anyone is able to download and collect data on projects.
As shown in Table \ref{tab:RQ2result}, we summarize and draw the top 3 popular projects using Gerrit tools.
We observe that for these CR papers, Qt project is the most studied project, with sixteen, fifteen, and twelve papers investigating Qt, OpenStack, and Android respectively.

\begin{figure}[pos = t]
\centering
\includegraphics[width=.8\textwidth]{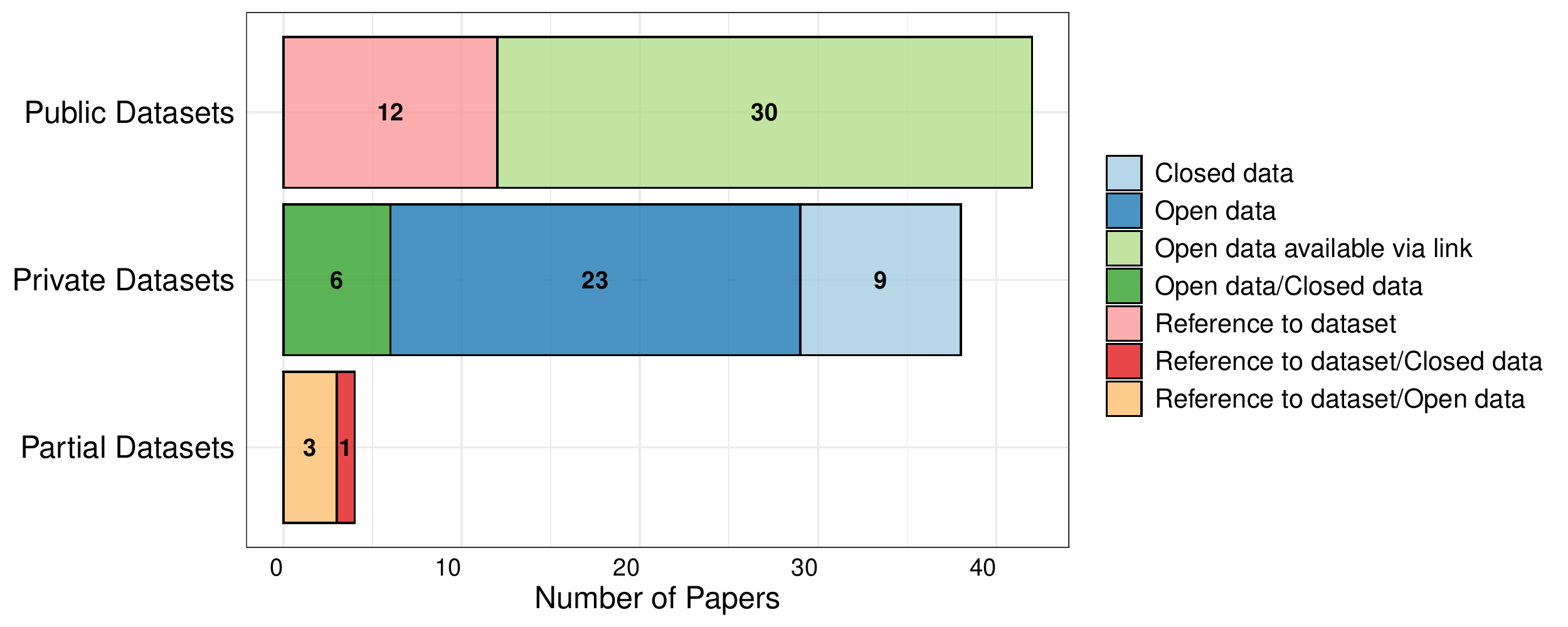}
\caption{Visual Map for RQ$_2$, showing the replicability of the collected papers. Note that papers analyzed in RQ2 are limited to quantitative and mixed-method papers. The figure shows that 42 papers (50\%) provide the public datasets.}
\label{fig:RQ2}
\end{figure}

Figure \ref{fig:RQ2} shows two important findings with the proportion of replicability.
The \textbf{first} finding is that there are in total 38 papers (i.e., 9 papers with closed data, 23 papers open data, and 6 papers open data/closed data) out of 84 papers (around 45\%) that do not provide any access to their datasets. 
Taking a closer look at the closed data, studies are usually conducted within industries, surveys and control studies.
An example of this paper is \citet{p19}, where the authors conducted research on how to reduce human efforts and improve review quality using the data from industry project named VMware.
For papers that labeled as open data, the researchers collected data from open source projects but did not share a replication package. For instance, \citet{p47} investigated whether or not pull requests encourage developers to upgrade out-of-date dependencies with the data from OSS (i.e., 7,470 projects in GitHub).
It could be argued that since the data is open source and available for anyone to download themselves.
The \textbf{second} finding is that, as shown in Figure \ref{fig:RQ2}, we observe that 42 papers out of 84 papers (around 50\%) released a replication package, either referred to a published dataset or released their own dataset via an online link.
For the work of \citet{p06}, authors referred to a dataset that was previously published~\cite{dataset_1} to revisit code ownership and its relationship with software quality.
Usually, papers release a link to the dataset. 
For instance, \citet{pj4} shared the dataset link (e.g., WebKit and Blinkin projects).
\respond{Upon a closer inspection on the dataset platforms of these replication package links, we found that GitHub/Bit Bucket and Personal/University are the most common dataset platforms (i.e., seventeen, respectively). While few researches make their dataset immutable (i.e., Permanent Storage), with three papers being classified.}
We summarize the available replication datasets with their URL links from these quantitative and mixed-method papers, referred to our Appendix (Section~\ref{appendix}).

%%%%%%%%%%%%%%%%%%%%%%%%%%%%%%%%%%%%%
%%%%%%%%%%%%%%%%%%%%%%%%%%%%%%%%%%

\begin{tcolorbox}
\textbf{Answering RQ2:}
    \respond{CR research not only relies on the data sources from the CR process but also largely uses the data sources from the software development process (i.e., issue tracking system and GitHub). We observed that 50\% of CR papers (i.e., 42 papers out of 84 papers) that use the quantitative method or mixed-method provide the public datasets (i.e., the replication links are provided in the papers).}
    \respond{The implication of RQ2 is that, to promote the validity of scientific findings, we encourage future researches to strive for a replicable dataset.}
\end{tcolorbox}

\subsection{\RqThree}

\begin{table}[pos = b]
\centering
\caption{Metric sets used in code review paper}
\begin{tabular}{lllllp{5cm}r}
\label{tab:metrics_kind}
Metric Set                                                                                                                                                                                                                                                                                                                                                & Product                   & Process                   & People                    & Others                    & Referred papers                                                                                                                                                                        & \# Related metrics \\ \hline
Experience   &                             & \checkmark & \checkmark &                           & \cite{p03, p10, p15, p21,p27, p28, p33, p49, p52, pj2, pj3,pj4, pj6, pj9, p60, p64, p67, p70, pj26,pj30, p73, p80,pj15} & 76    \\
Code                                                                                                                                           & \checkmark & \checkmark &                           &                           & \cite{p49,p67,p60,p03,p27,pj2,p06,p10,p23,p28,p33,pj5,pj6,pj9,p57,p64,p70,pj3,pj28,p15,p62,p52,pj30,p80,p29,pj4,p16,p04}                                                                                                                                                              & 72    \\
Ownership                                                                                                                                                                                                                                       &                           &                           & \checkmark &                           &  \cite{p67,p27,pj2,p06,p10,p23,p28,p33,pj5,p64,p70,pj3,pj15,p21,p29,p04, p57}                                                                                                                                                                                     & 38    \\
Comment                                                                                                                                 &                           & \checkmark &                           &                           &     \cite{p49,p67,p03,p27,p10,p23,p33,pj6,p57,p64,p70,pj3,p15,p21,p52,p16, pj28}                                                                                                                                                                                  & 37    \\
File                                                                                                                                                                                                                                                                                                                   & \checkmark & \checkmark &                           &                           &                                                                                                       \cite{p49,p67,p03,p27,pj2,p06,p10,p28,p33,pj6,pj9,p57,p70,pj3,pj15,p15,p52,pj30,p80,p29,p16}                                                                                & 31    \\
Participant                                                                                                                                                                                                                                                    &                           &                           & \checkmark &                           &                                        \cite{p03,p27,p06,p10,p23,p28,pj6,pj9,p64,pj28,pj15,p62,pj30,p80,p29,p16,p33,p15,p57}                                                                                                                                               & 31    \\
Temporal                                                                                                                                                                                                                                                          &                           & \checkmark &                           &                           &                                                                                                                                \cite{p49,p67,p27,p10,p23,pj6,pj9,p57,p64,p70,pj3,p62,p73}                                                       & 26    \\
Revision                                                                                                                                                                                                                                                   & \checkmark & \checkmark &                           &                           &                                                                                                                                                            \cite{p67,p49,p03,p27,pj2,pj6,pj9,p23,p28,p64,p70,pj15,p62,p15,p80,p52,pj30}                           & 24    \\
Description                                                                                                                                                                                                                                                         & \checkmark & \checkmark &                           &                           &                                                                                                                                                  \cite{p49,pj2,pj9,pj3,p80,p29,p67,p70,pj30}                                     & 21    \\
Module                                                                                                                                                                                                                                                                                            & \checkmark &                           &                           &                           &     \cite{p67, p60, p27, pj2, p33, pj6, pj9, p70, pj3,p15,pj30,p80,p29, pj4, p16, p04} & 21    \\
Defect                                                                                                                                                                                                                                                                                                              &                           & \checkmark &                           &                           & \cite{pj3, pj4, pj5, p57, p60, p62}                                                                                                                                  & 14    \\
Queue                                                                                                                                                                                                                                                                                                                           &                           & \checkmark &                           &                           & \cite{p49, p60, p62, p80,pj4,p33}                                                                                                                                            & 6     \\
Workload                                                                                                                                                                                                                                                                        &                           &                           & \checkmark &                           & \cite{pj28, p49, pj3}                                                                                                                                      & 5     \\
Decision                                                                                                                                                                                                                                                                                                                 &                           & \checkmark &                           &                           & \cite{p52, p80, p27,pj15}                                                                                                                                                 & 5     \\
Language                                                                                                                                                                                                                                                                                                         & \checkmark &                           &                           &                           & \cite{pj2, p80}                                                                                                                                                      & 4     \\
Log                                                                                                                                                                                                                                                                                                                                     &                           & \checkmark &                           &                           & \cite{p49}                                                                                                                                                           & 1     \\ \hline
Email                                                                                                                                                                                                                                                                                                                           &                           &                           &                           & \checkmark &                                                                        \cite{p27}                                                                                                               & 11    \\
Collaboration                                                                                                                                                                                                                                             &                           &                           &                           & \checkmark &                                                                                                                                              \cite{p16,pj2}                                         & 10     \\
Build related                                                                                                                                                                                                                                                                                                        &                           &                           &                           & \checkmark &                                                                                   \cite{p67}                                                                                                    & 4     \\
Project                                                                                                                                                                                                                                                                                                         &                           &                           &                           & \checkmark &                                                                                      \cite{p16,pj4}                                                                                                 & 9     \\
Others                                                                                                                                                                                                                                                                                 &                           &                           &                           & \checkmark &                                                                                                                 \cite{p67,p60,p03,p33,p64,pj28,pj15,p80,p73}                                                                     & 11     \\ \hline
Total                                                                                                                                                                                                                                                                                                                                                         &                           &                           &                           &                           & \multicolumn{1}{r}{31}                                                                                                                                                                & 457  
\end{tabular}
\end{table}

Table~\ref{tab:metrics_kind} shows sixteen core metrics sets based on 457 identified metrics with their corresponding review aspect and frequency.
We summarize three findings.
\textbf{First}, we observe that \texttt{Experience} and \texttt{Code} are the two most frequently used metric sets, which are far more than other classified metric sets.
In detail, 76 and 72 metrics are involve in \texttt{Experience} and \texttt{Code} sets respectively.
\texttt{Experience} is referred to those metrics computing the patch author or reviewer experience in submitting historical patches.
Given an example, \citet{pj28} took four experience related metrics into account: \texttt{Reviewer Code Authoring Experience}, \texttt{Reviewer Reviewing Experience}, \texttt{Patch Author Code Authoring Experience}, \texttt{Patch Author Reviewing Experience}.
\texttt{Code} denotes to those metrics that focus on measuring the source code of a system.
For instance, \citet{pj6} computed the number of added and deleted code lines.
Apart from the general patch size, \citet{p80} proposed in-depth code related metrics such as \texttt{Class churn}, \texttt{Loop churn}, and \texttt{Method churn}.
\respond{Others refers to those metric sets that can not fit to the common metric sets with their definition.}
\textbf{Second}, from the view of the referred paper number, the result indicates that \texttt{Code} is considered as the first ranking basic metric set applied to model constructions.
28 out of total 31 papers (i.e., around 90\%) take such metric set in to account.
The \textbf{third} observation is that as we see from Table~\ref{tab:metrics_kind}, one metric set can represent multiple CR aspects.
For example, \texttt{File} metric set can either belong to \texttt{Product} and \texttt{Process}.
Those metrics that compute the number of added and deleted files in the patch are regarded as \texttt{Process} aspect~\cite{pj2,pj9,p15,p52}.
On the other hand, researchers calculate the file entropy~\cite{pj3,pj6,pj9}.
Such metrics are more likely to be dynamic and are viewed as \texttt{Process} aspect.
To answer the correlation between metrics and their corresponding research topics, we summarize nine topics regarding code review and list related papers below:
\begin{itemize}
    \item \textit{Quality Assurance}: refers to papers focusing on the code quality such as bug fixing~\cite{p06,p10,p23,p28,p33,p57, p64,p70,pj5,pj6,pj9} 
    \item \textit{Acceptance Predication}: refers to papers focusing on predicting the decision of the patches~\cite{p03,p27,pj2} .
    \item \textit{Review Process}: refers to papers exploring or comparing the different peer review models~\cite{p15,p62}.
    \item \textit{Review Participation}: refers to papers focusing on the reviewer participation~\cite{pj3,pj15,pj28}.
    \item \textit{Review Process Prediction}: refers to papers focusing on predicting the period taken to complete the review~\cite{p52,p80,pj30,p29}.
    \item \textit{Review Process Comments}: refers to papers focusing on predicting usefulness of review comments ~\cite{p21}.
    \item \textit{CI \& Review}: refers to papers focusing on the correlation between CI implementation and code review~\cite{p49,p67}.
    \item \textit{Test \& Review}: refers to papers focusing on the correlation between test and code review~\cite{p04,p73}.
    \item \textit{Technical/Non-Technical \& Review} refers to papers investigating the technical and non-technical factor impact on the code review~\cite{pj4,p16,p60}.   
\end{itemize}

%%%%%%%%%%%%%%%%%%%%%%%%%%%%%%%%%%%%%
\begin{figure}[]
\centering
\includegraphics[width=.9\textwidth]{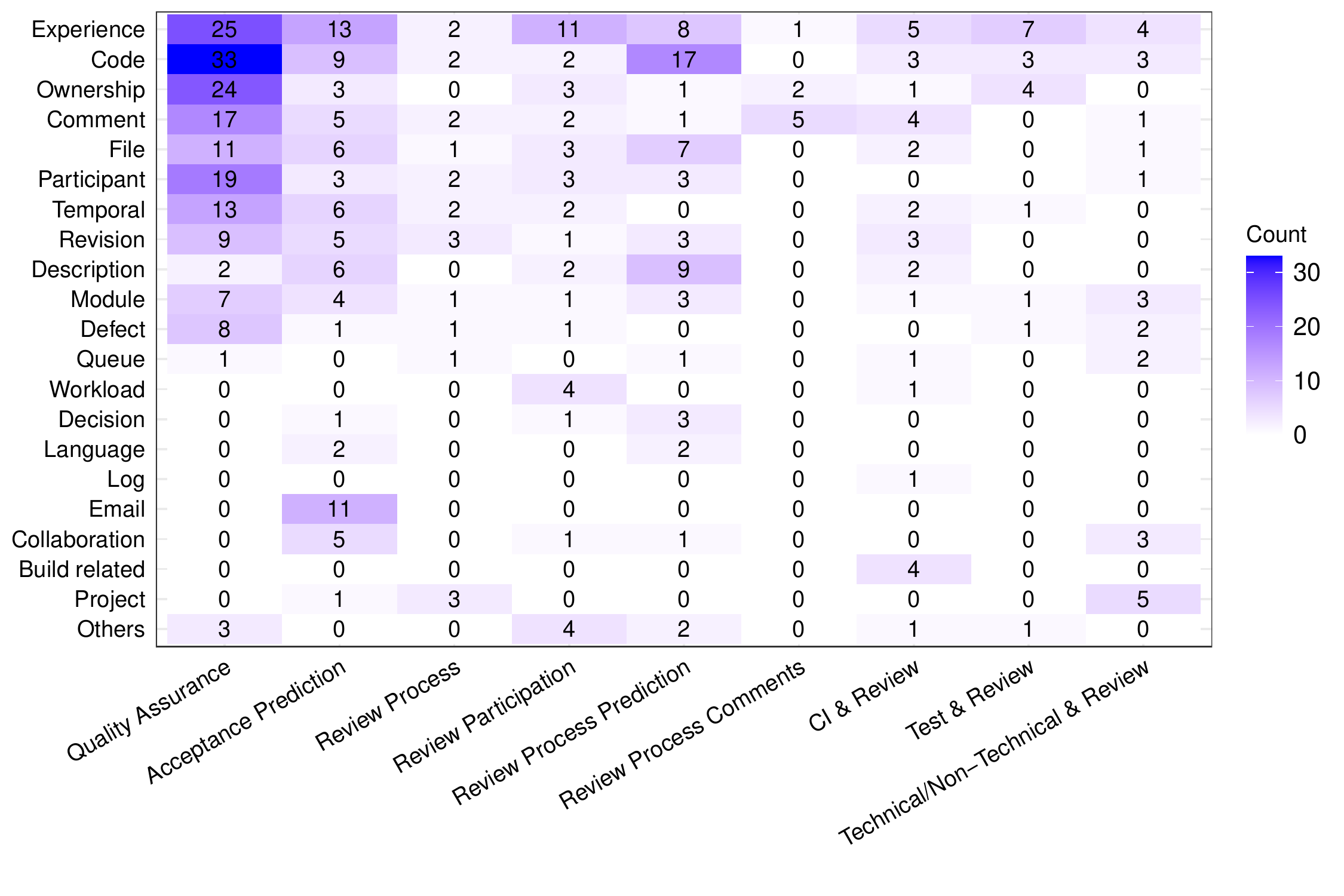}
\caption{Nine research topics with their target metric sets. The figure shows that different research topics tend to target particular metric sets.}
\label{fig:heat_map}
\end{figure}
%%%%%%%%%%%%%%%%%%%%%%%%%%%%%%%%%%%%%

Figure~\ref{fig:heat_map} maps the metric sets with nine CR research topics. Two findings are observed based on the related paper list and Figure~\ref{fig:heat_map}.
\textbf{First}, we find that \textit{Quality Assurance} related topics are more likely to use CR metrics to conduct the research, i.e., eleven papers are identified within this topic. 
For instance, \citet{pj5} conducted an empirical study to investigate the relationship between post-release defects and code review practices such as coverage, participation, and reviewer expertise.
Their findings confirmed the intuition that poorly-reviewed code has a negative impact on software quality.
The second popular topic is \textit{Reviewer Process Prediction}, i.e., four papers retrieved, which refers to papers focusing on predicting the review period.
One example of the process prediction topic is that in the study of \citet{pj30}, they applied a couple of eighteen metrics like source code, textual information, experience, and social connection related metrics to recommend pull requests that can be quickly reviewed by reviewers. 
The \textbf{second} finding is that different research topics use particular metric sets.
As shown in Figure~\ref{fig:heat_map}, \textit{Review Process Comments} only adopt \texttt{Experience}, {Code}, and \texttt{Comment} into their statistic model construction. 
However, for \textit{Quality Assurance}, \textit{Acceptance Predication}, and \textit{Review Participation},  the metric sets are diverse considering comprehensive angles.
For instance, the \textit{Review Participation} topic takes almost all kinds of metric sets into accounts except for \texttt{language}, \texttt{Queue}, and \texttt{Log}.
In addition, for those specific research topic, the particular metric sets will be computed specially such as \texttt{Build related} metric sets used in the \textit{CI \& Review} topic.
The detailed metrics within each metric set are listed in the Appendix (Section~\ref{appendix}).

\begin{tcolorbox}
    \textbf{Answering RQ3:}
    \respond{Sixteen core metrics sets are grouped based on 457 metrics extracted from the quantitative papers, and nine research topics that use these metrics are classified (quality assurance, review process prediction, acceptance prediction, and so on). We observe that the SE topic of quality assurance is more likely to use CR metrics to conduct the research, i.e., eleven papers. In addition, experience and code are the two most frequently used metric sets.} \respond{From the mapping between metric sets and research topics, we find that it has the potential to benchmark the metric based CR research, as different research topics tend to use particular metric sets.}
    \respond{The implcation of RQ3 is that, we encourage the researchers to take the existing metrics into account when conducting a certain topic.}
\end{tcolorbox}

\section{Comparative Analysis}
\respond{In this section, we followed the protocol of comparative analysis provided by the work of \citet{Feldt_08} to compare and highlight the novelty of our work against existing systematic reviews. The systematic review studies were identified using the following search string: \textit{``systematic review'' AND ``code review''} and by searching a broad collection source (i.e., ACM Digital Library, IEEE Xplore, Science Direct, and SpringerLink databases). We exclude papers that did not explicitly state in the title or abstract that they were systematic reviews or were not published in the traditional SE domain based venues. The search results in a total of three systematic reviews~\citep{2008_mapping,short_mapping,Iwor} before or in 2019.}

\begin{table}[pos = b]
\centering
\caption{Existing Systematic Review Characteristics}
\label{comparative_analysis}
\begin{tabular}{lrrr}
Reference Systematic Reviews& \citet{2008_mapping}& \citet{short_mapping}& \citet{Iwor} \\ \hline
\textbf{Research Goals}                      &   &   &   \\ \hline
Identify Best and Typical Practices & \checkmark  &  \checkmark &   \\
Classification and Taxonomy         & \checkmark  &   &  \checkmark \\
Emphasis on Topic Categories        &  \checkmark &  \checkmark &   \\
Identify Publication Fora           &  \checkmark &  \checkmark &  \checkmark \\ \hline
\textbf{Inclusion Requirements}              &   &   &   \\ \hline
Research is Within Focus Area       &  \checkmark &  \checkmark &  \checkmark \\
Empirical Methods Used              &  \checkmark &  \checkmark &  \checkmark \\ \hline
\textbf{Number of Included Articles}         &   &   &   \\ \hline
Potentially Relevant Studies        &  229 & 873  &  - \\
Relevant Studies (Included)         & 153 (1980-2008)  & 177 (2005-2018)  & 13 (2007-2018)  \\ \hline
\textbf{Means of Analysis}                   &   &   &   \\ \hline
Meta Study                          &  \checkmark &  \checkmark &  \checkmark \\
Comparative Analysis                &   &   &   \\
Thematic Analysis                   &  \checkmark &  \checkmark &   \\
Narrative Summary                   &  \checkmark &   &  
\end{tabular}
\end{table}

\respond{For each of the three CR systematic reviews, we characterize them based on their research goals, criteria for inclusion requirements, the number of papers, and means of analysis. Table~\ref{comparative_analysis} shows the characteristics of the three existing systematic reviews. Below, we now discuss the differences between our study and the existing systematic reviews in detail, with the aspect of research goals, the process, and the breadth and depth: }
\respond{\begin{itemize}
    \item \textbf{Difference in Research Goals:} The systematic review conducted by \citet{2008_mapping} aims at reviewing how the software inspection field has evolved between 1980 and 2008. Their focus is set in the context of the software inspection. However, our mapping study addresses the contemporary tool-based code review, which is a light variant of the software inspection and has been widely adopted in both industrial and open-source projects. Although \citet{short_mapping} and \citet{Iwor} conducted the systematic review related to contemporary tool-based code review, the goal of our mapping study is also different from them. The main goal of the work by \citet{short_mapping} is to observe the evolution of the research topic, while the work by \citet{Iwor} is to gather evidence on the extent of the work related to refactoring-awareness during code review. Differently, our study aims at investigating the potential of benchmarking in the aspect of datasets and metrics. We believe that a common benchmark could facilitate future CR related researches and help researchers to propose new approaches and compare against existing ones.
    \item \textbf{Difference in Process:} We observe two main differences in the process when compared to the two systematic reviews related to the tool-based review. First, compared to the work of \citet{Iwor}, we include the thematic analysis (i.e., classification schema of methodologies, contributions, data, and metrics). For the systematic mapping study, thematic analysis is an interesting analysis method, which helps to see which categories are well covered in terms of number of publications~\citep{Feldt_08}. Second, compared to both work by \citet{Iwor} and \citet{short_mapping}, we conduct an in-depth narrative summary analysis with qualitative
    review of each paper, as both of them are served as a preliminary study. 
    \item \textbf{Difference in Breadth and Depth:} Two differences are summarized, based on the systematic reviews related to the tool-based review. On the one hand, \citet{Iwor} focused on the specific theme of tool-based code review, i.e., refactoring-awareness. However, our study covers all potential themes. On the other hand, in the study of \citet{short_mapping}, they extracted CR related papers from all possible venues (i.e., 177 papers are retrieved between 2005 and 2018). While to ensure the paper quality and form a best representative view for future researches, we only focus on the papers that are published in the premium venues (i.e., 112 papers are retrieved between 2011 and 2019).
\end{itemize}}

\section{Towards a Common Benchmark of Dataset and Metric}

\begin{table}[pos = b]
\caption{A summary of common metric sets and datasets used in various SE topics.}
\label{tab:discussion}
\resizebox{\textwidth}{!}{
\begin{tabular}{@{}lrp{6cm}rp{3.5cm}@{}}
CR Topic & \# Papers & Common Metric Sets                                                                    & Common Datasets & Review Settings                        \\ \midrule
Quality Assurance       & 11        & Code (100\%), Ownership (82\%), File, Participant (73\%), Experience (64\%)       & 18\%            & Gerrit (8), Pull-based (0), Others (3) \\
Review Process Prediction      & 4         & Code (100\%), Participant, File, Module, Description, Experience, Revision (75\%) & 0               & Gerrit (1), Pull-based (3), Others (0) \\
Acceptance Prediction       & 3         & Code, Experience, File (100\%), Comment, Ownership (66\%)                         & 0               & Gerrit (1), Pull-based (1), Others (1) \\
Review Participation     & 3         & Experience (100\%), Code, File, Workload, Participant, Comment (66\%)                                   & 0               & Gerrit (2), Pull-based (1), Others (0) \\
Review Process     & 2         & Code, Revision, Participant (100\%)                                               & 0               & Gerrit (0), Pull-based (1), Others (1) \\
CI \& Review      & 2         & Code, Comment, Description, Experience, File, Revision, Temporal (100\%)          & 0               & Gerrit (0), Pull-based (2), Others (0) \\
Test \& Review      & 2         & -                                                                                 & 0               & Gerrit (1), Pull-based (0), Others (1) \\
Technical/Non-Technical \& Review     & 3         & Code, Module (100\%), Defect, Experience, Defect (66\%)                           & 67\%            & Gerrit (0), Pull-based (1), Others (2) \\ \bottomrule
\end{tabular}}
\end{table}

Although our results suggest that CR research is mostly driven by empirical evaluation, we conclude that at this stage, we cannot benchmark CR studies.
However, the existing datasets and metrics do show potential for creating a benchmark.
With the rise of machine learning and AI techniques, CR researchers will soon need to agree on the common set of metrics that should be included to accurately compare such techniques against each other.
Having a benchmark will facilitate new researchers, including experts from other fields, to innovate new techniques and build on top of already established methodologies.
This mapping commonalities between metrics and datasets is shown in Table \ref{tab:discussion}. 

Table \ref{tab:discussion} suggests that there exists a regular group of metric set combinations commonly used for papers that are addressing a specific SE topic.
We selected metrics that are commonly mentioned in more than 60\% of the classified papers (i.e., from RQ3). 
For instance, in \textit{Quality Assurance} related papers, \textit{Code} is computed for all (i.e., 100\%) and around 82\% of papers take \textit{Ownership} into account.
In \textit{Acceptance Prediction} related papers, all three papers compute the metrics of \textit{Code}, \textit{File}, and \textit{Experience}.
On the other hand, Table \ref{tab:discussion} also shows that common datasets were not commonly adopted by researchers.
For instance, since two of the three papers in \textit{Technical/Non-Technical and Review} were written by the same authors, the ratio for having a common dataset is high (i.e., 67\%).
This means that researchers from different groups prefer to construct their own datasets to conduct their study.
Another reason is because the technology used to generate datasets constantly evolve, thus, deeming any prior datasets as being outdated.
Furthermore, since more datasets are taken from either the Gerrit or GitHub Pull-request API, they sometimes miss the essential elements needed for a specific study.
This process can be time-consuming, and could be easily resolved by using a benchmark.
We also find that different review settings (i.e., Gerrit and Pull-based) have different emphases on the SE topics.
For example, \textit{Quality Assurance} datasets are almost from Gerrit while in \textit{CI and Review} datasets are all from Pull-based review settings.
One possible reason is that some specific metric sets are not easily available to be retrieved from different review settings.

\section{Threats To Validity}
\respond{We now discuss threats to the validity of our mapping study.}

\respond{\textbf{External validity.} External validity is concerned with our ability to generalize based on our results. The results of this mapping study are considered with regard to the CR domain,
while the validity of conclusions is applicable only to the CR context.
The external validity threats are thus not applicable.
}

\respond{\textbf{Construct validity.} Construct validity is concerned with the degree to which our measurements capture what we aim to study. 
During the qualitative analysis, especially for the methodology and the contribution classification, methodologies and contributions may be miscoded due to the subjective nature of our coding approach. To mitigate this threat, three co-authors sat in a round-table and did the classification. If a dispute occurred, the full contents of the papers were discussed before the consensus was reached.}

\respond{\textbf{Internal validity.} Internal validity is the approximate truth about inferences regarding cause-effect or causal relationships.  
We summarize three potential internal threats. The first threat is related to the venue selection. In this mapping study, we only consider 34 top venues considering their online citation indices and feedback from the software engineering community, similar to the work of~\citet{topic_tse}. Thus, there will always be a venue missing from such a study and can be considered. However, we believe these 34 top venues can represent the best practice for the CR research.}
\respond{The second threat is related to the paper selection of the studies during the screening process. 
Due to the large amount of hits from our search string, our initial step includes the first author scanning through and discarding papers based on titles and abstracts, which potentially raises a bias in the paper selection.
Nevertheless, we are confident of this threat, as the first author is an existing code review researcher and is familiar with the domain.}
\respond{The third possible internal threat is with regard to the terms that are used in our search string. The case might exist that the search string will not cover all terms. 
To reduce such risk, in the initial round, we manually checked twenty CR related papers to group the term candidates, and we were confident that the existing search term is sufficient.
}

\respond{\textbf{Conclusion validity.} Conclusion validity is the degree to which conclusions we reach about relationships in our data are reasonable.
In the case of our datasets, there is a threat that our grouping is not accurate.
Since there is no related work that has similar results, we cannot verify our findings.
To mitigate this, we rely on the systematic guidelines for our outcomes. 
In addition, we publish a website and open it to both researchers and practitioners to criticize or add to our results.}

\section{Conclusion}
\respond{Code review (CR), as a well-known practice, plays a vital role in software quality assurance. In the recent decade, the contemporary review tools have made the review process now being light-weight and have been widely adopted in both open-source and industrial projects. Due to the availability of datasets brought by these review tools, CR related researches are largely carried out. In order to understand the state-of-the-art practices,, we first conduct a systematic mapping study to provide a visual summary of the benchmark potential within the CR domain through 112 papers that are published in premium conferences and journals.}

\respond{Three main maps are visualized. First, concerning the map between methodologies and contributions, we find that 65\% of CR researches use sound evaluation methodology (i.e., 73 papers), targeting particularly socio-technical and understanding of CR. While, there is a lack of papers that report the experience and propose solutions to deal with CR problems (i.e., four papers and thirteen papers, separately). Second, concerning the map of the datasets, our results show that few researches provide replicable datasets, i.e., around 50\%. Third, concerning the map of metrics, we identify 457 metrics which are grouped into sixteen core metrics sets, and we observe that the SE topic of quality assurance is more likely to use CR metrics to conduct the research, with eleven papers being classified.
Additionally, we find that different research topics use particular metric sets, which provides the potential for a benchmark.}

\respond{The next step is the creation of a benchmark to facilitate future research against the state-of-the-art.
With the rise of machine learning and AI techniques, a common benchmark is needed as it will facilitate CR researchers to accurately compare techniques against each other and propose new approaches. To encourage this framework construction, we provide a listing of methodologies and contribution, 42 public datasets, and 457 identified metrics across nine research topics which is available at \respond{\url{https://naist-se.github.io/code-review/}}.}

\section*{Acknowledgment}
This work has been supported by JSPS KAKENHI Grant Numbers 18H04094 and 20K19774 and 20H05706.

\bibliography{cas-refs}

\section{Appendix: Metric Sets and Available Datasets.}
\begin{appendices}
\section{metrics set listing with sample keywords}
%%%%%%%%%%%%%%%%%%%%%%%%%%%%%%%%%%%%%%%
% Following shows the keywords of each group.
{\footnotesize
\begin{itemize}
    \item EXPERIENCE: author/reviewer/recent/library/java/author accept experience (\#Patches, \#PRs, \#Commits), proportion of changes\\ without discussion/hastily reviewed changes/self-approved changes,  age of the PR author,  core member, followers, \\ hours, reviewed churn/commit, subsystem changes/merge\_ratio
    \item CODE: size, added/deleted LOC, class/test/method/loop/reference/relative/src/comulative/conditional statemants churn,\\ complexity, segs added/deleted/updated,  chunk in/out
    \item COMMENT: \#/avg./author/reviewer/in-code/no comments, comment length/sentiment/, hastily, code element/\\question/word ratio, reading ease, conceptual similarity, review disagreement 
    \item OWNERSHIP:\#/major/minor/authors, author/code/code review ownership, Top RSO/TCO, self approval/verify, file\_dev\_num 
    \item FILE:\# files, added/deleted files, entropy 
    \item PARTICIPANT:\#reviewers/developers, participation, proportion of major/minor author (reviewer), interaction
    \item TEMPORAL:age, review/integrate/issue fix/merge/time, discussion/response lag, review window, duration
    \item REVISION:\#revisions, \#iterations, \#commits, version, \#file changes, proportion of reviewed changes/churn
    \item DESCRIPTION:description/title length, purpose, has bug/document/feature/refactor/improve/, readability
    \item MODULE:\#modified directories/subsystems/path, component, test code
    \item DEFECT:\#prior defects, priority/severity bug   
    \item WORKLOAD:overall/directory/merge workload, review queue, \#concurrent/remaining reviews
    \item DECISION:merge/accept/deprecate change or not
    \item LANGUAGE:file types, language num, setting or config.
    \item QUEUE:active, review/project queue  
    \item LOG:stack trace attached 
\end{itemize}

\begin{itemize}
    \item EMAIL: CC, thread, \#email, mail 
    \item COLLABORATION:degree/closeness/betweenness/eigenvector centrality, clustering\_coefficient, familiarity, social distance, collaborator
    \item BUILD RELATED:failed/PR builds, first/last build status
    \item PROJECT: organization, team, affiliation, external, maturity    
    \item OTHERS:i.e., day of the week, business hours, Intra-Branch, right\_venue, etc.
\end{itemize}

}

\section {available datasets}
\label{appendix}
\begin{tabularx}{\textwidth}{@{}lX@{}l@{}lX@{}lX}
\caption{Listing of fully replicated papers with their available datasets}\\
\label{tab:appendix}
\textbf{Reference} & \textbf{Available dataset} \\\hline\hline
\citet{p03} & \url{https://cs.uwaterloo.ca/~okononen/shopify}\\
\citet{p04}& \url{https://doi.org/10.5281/zenodo.1172419} \\
\citet{p05} &  \url{http://dijkstra.cs.virginia.edu/fmri/}\\
\citet{p06} & Who does what during a
Code Review? An extraction of an OSS Peer Review
Repository~\cite{dataset_1}\\
\citet{p12} & \url{https://sites.google.com/a/utexas.edu/critics/}\\
\citet{p17} & \url{https://github.com/Mining-multiple-reposdata/experimental dataset}\\
\citet{p21} & \url{http://homepage.usask.ca/masud.rahman/revhelper}\\
\citet{p23} & Who does what during a
Code Review? An extraction of an OSS Peer Review
Repository~\cite{dataset_1} / The Impact of Code Review Coverage and Code Review
Participation on Software Quality~\cite{p28}\\
\citet{p28} & \url{http://sailhome.cs.queensu.ca/replication/reviewing\_quality/}\\
% \citet{p29} & \\
\citet{p31} & Mining the
modern code review repositories: A dataset of people, process and
product~\cite{xinyang_2016}\\
\citet{p32} &  \url{https://github.com/yiu31802/icsme2016}\\
\citet{p33} & \url{https://cs.uwaterloo.ca/~okononen/bugzilla\_public\_db.zip}\\
\citet{p34} & Who should review my code?~\cite{p65}\\
\citet{p35} & \url{http://tinyurl.com/kzw43n6}\\
\citet{p40} & \url{http://amiangshu.com/VCC/index.html}\\
\citet{p42} & Gerrit software code review data from Android~\cite{android_msr2013}\\
\citet{p46} & \url{https://mhepaixao.github.io/architecture\_awareness/}\\
\citet{p47} & \url{https://github.com/alt-code/Research/tree/master/VersionBot}\\
\citet{p48} & \url{https://www.uni-due.de/~hw0433/}\\
\citet{p49} &\url{https://prdeliverydelay.GitHub.io/\#datasets}\\
\citet{p52} & \url{http://cstar.whu.edu.cn/p/cpr/}\\
\citet{p58} & \url{https://github.com/software-rebels/DesignInCodeReviews-ESEM2018}\\
\citet{p60} & \url{https://cs.uwaterloo.ca/~obaysal/webkit\_data.sqlite}\\
\citet{p61} & \url{http://tinyurl.com/pull-resourcesreplication}\\
\citet{p62} & \url{https://bitbucket.org/foundjem/ml-issuetk/src}\\
\citet{p65} & \url{http://github.com/patanamon/revfinder}\\
\citet{p66} & \url{http://ser.soccerlab.polymtl.ca/ser-repos/public/tr-data/2015-saner-code-reviews.zip}\\
\citet{p67} & \url{https://goo.gl/KjTpxp}\\
\citet{p68} &  \url{https://github.com/felipeebert/confusion-code-reviews}\\
\citet{p70} & \url{http://gsyc.es/~jgb/repro/2015-msr-grimoire-data}\\
\citet{p72} & \url{https://zenodo.org/record/3354510\#.X1i3W2czbLA}\\
\citet{p73} & \url{https://www.mediafire.com/folder/3b5ey849y9flx/Test-Driven\_Code\_Review\_-\_Online\_Appendix}\\
\citet{p76} & \url{https://tinyurl.com/y3yk6oey}, \url{https://github.com/Tbabm/PRSummarizer}\\
\citet{p79} & \url{https://github.com/software-rebels/ReviewLinkageGraph}\\
\citet{pj3} & \url{http://sailhome.cs.queensu.ca/replication/review\_participation/}\\
\citet{pj4} & \url{https://cs.uwaterloo.ca/~obaysal/webkit\_data.sqlite}\\
\citet{pj5} & \url{https://sailhome.cs.queensu.ca/replication/reviewing\_quality\_ext/}\\
\citet{pj6} & \url{https://github.com/software-rebels/JITMovingTarget}\\
\citet{pj9} & \url{http://research.cs.queensu.ca/~kamei/jittse/jit.zip}\\
\citet{pj13} & \url{https://github.com/yuyue/pullreq_ci}\\
\citet{pj27} & Mining the
modern code review repositories: A dataset of people, process and
product~\cite{xinyang_2016}\\
\citet{pj28} & \url{https://github.com/sruangwan/replication-human-factors-code-review/}\\
\hline
\end{tabularx}

\end{appendices}

\end{document}